\documentclass[11pt, oneside]{article}

\usepackage[english]{babel}
\usepackage{geometry}  
\usepackage{graphicx}      
\usepackage{tabularx}      
\usepackage{amssymb}
\usepackage{amsmath}

\usepackage{xltabular}
\usepackage[hyphens]{url}
\usepackage[hidelinks]{hyperref}
\hypersetup{breaklinks=true}

\usepackage{mdframed}
\usepackage[backend=biber,style=chem-rsc, doi=false,isbn=false, date=year, url=false, autopunct=true]{biblatex}
\makeatletter
\renewbibmacro*{textcite}{%
  \iffieldequals{namehash}{\cbx@lasthash}
    {\usebibmacro{cite:comp}}
    {\usebibmacro{cite:dump}%
     \ifbool{cbx:parens}
       {\bibclosebracket\global\boolfalse{cbx:parens}}
       {}%
     \iffirstcitekey
       {}
       {\textcitedelim}%
     \usebibmacro{cite:init}%
     \ifnameundef{labelname}
       {\printfield[citetitle]{labeltitle}}
       {\printnames{labelname}}%
     \addspace
     \ifnumequal{\value{citecount}}{1}
       {\usebibmacro{prenote}}
       {}%
     \mkbibsuperscript{\usebibmacro{cite:comp}}%
     \stepcounter{textcitecount}%
     \savefield{namehash}{\cbx@lasthash}}}
\makeatother

\usepackage{csquotes}
\usepackage{cleveref}
\usepackage{siunitx}
\usepackage{geometry}
\usepackage{multirow}
\usepackage[dvipsnames,table,xcdraw]{xcolor}
\definecolor{tumblue}{rgb}
{0.6431372549,0.2196078431,0.1882352941}

\usepackage{xspace}
\usepackage{caption}
\usepackage{subcaption}
\usepackage{authblk}

\usepackage{marvosym}

\usepackage{sectsty} 

\sectionfont{\color{tumblue}\selectfont\sffamily}
\subsectionfont{\color{tumblue}\selectfont\sffamily}
\subsubsectionfont{\color{tumblue}\selectfont\sffamily}
\paragraphfont{\selectfont\sffamily}

\IfFileExists{./fonts/Heliotrope-Bold.otf}{
    \usepackage{fontspec}
    \setmainfont{Heliotrope}[Extension = .otf, UprightFont = *-Regular, ItalicFont = *-Italic, BoldFont=*-Bold, Path = ./fonts/]
}{
    \usepackage{biolinum}
    
}

\usepackage{lipsum}
\usepackage[labelfont=bf]{caption}
\usepackage[nonumberlist,acronym]{glossaries}
\setacronymstyle{long-short}
\newglossaryentry{dft}{
    type=\acronymtype, 
    name={DFT}, 
    description={density functional theory},
}

\newglossaryentry{mae}{
    type=\acronymtype, 
    name={MAE}, 
    description={mean absolute error},
}

\newglossaryentry{rmse}{
    type=\acronymtype, 
    name={RMSE}, 
    description={root mean squared error},
}

\newglossaryentry{llm}{
    type=\acronymtype, 
    name={LLM}, 
    description={large language model},
}

\newglossaryentry{ml}{
    type=\acronymtype, 
    name={ML}, 
    description={machine learning},
}

\newglossaryentry{mof}{
    type=\acronymtype, 
    name={MOF}, 
    description={metal-organic framework},
}

\newglossaryentry{casp}{
    type=\acronymtype, 
    name={CASP}, 
    description={Critical Assessment of Structure Prediction},
}

\newglossaryentry{harking}{
    type=\acronymtype, 
    name={HARKing}, 
    description={Hypothesizing After the Results are Known},
}

\newglossaryentry{api}{
    type=\acronymtype, 
    name={API}, 
    description={application programming interface},
}

\newglossaryentry{tco}{
    type=\acronymtype, 
    name={TCO}, 
    description={transparent conductive oxide},
}

\newglossaryentry{mlff}{
    type=\acronymtype, 
    name={MLFF}, 
    description={machine learning force field},
}

\newglossaryentry{md}{
    type=\acronymtype, 
    name={MD}, 
    description={molecular dynamics},
}

\newglossaryentry{roc}{
    type=\acronymtype, 
    name={ROC}, 
    description={receiver operating characteristic},
}

\newglossaryentry{pr}{
    type=\acronymtype, 
    name={PR}, 
    description={precision recall},
}

\newglossaryentry{mcq}{
    type=\acronymtype, 
    name={MCQ}, 
    description={multiple choice question},
}

\newglossaryentry{pbe}{
    type=\acronymtype, 
    name={PBE}, 
    description={Perdew–Burke–Ernzerhof},
}

\newglossaryentry{gnn}{
    type=\acronymtype, 
    name={GNN}, 
    description={graph neural network},
}

\newglossaryentry{pxrd}{
    type=\acronymtype, 
    name={PXRD}, 
    description={powder X-ray diffraction},
}

\newglossaryentry{scxrd}{
    type=\acronymtype, 
    name={SC-XRD}, 
    description={single crystal X-ray diffraction},
}
\makeglossaries

 \usepackage{microtype}
 \usepackage{fontawesome}

\definecolor{linkcolor}{RGB}{0, 0, 0}

\usepackage{credits}
\usepackage{orcidlink}
\usepackage{hyperref}
\usepackage{multirow}
\usepackage{threeparttable}
\usepackage{arydshln}
\usepackage{xcolor}
\usepackage{booktabs}
\usepackage[most]{tcolorbox}

\definecolor{implicationbg}{RGB}{245, 245, 245}
\definecolor{examplebg}{HTML}{C3E1F0}
\definecolor{darkgreen}{HTML}{92C4DE}
\newtcolorbox{practicalbox}[1][Practical Implications]{%
  enhanced,
  breakable,
  colback=implicationbg,
  colframe=tumblue,
  fonttitle=\bfseries,
  coltitle=tumblue,
  attach boxed title to top left={yshift=-\tcboxedtitleheight/2, xshift=1em},
  boxed title style={size=small, colback=white, colframe=tumblue, sharp corners},
  top=12pt,
  bottom=6pt,
  left=8pt,
  right=8pt,
  title=#1,
}

\newtcolorbox{examplebox}[1][Example]{%
  enhanced,
  breakable,
  colback=examplebg!30,
  colframe=darkgreen,
  fonttitle=\bfseries,
  coltitle=darkgreen,
  attach boxed title to top left={yshift=-\tcboxedtitleheight/2, xshift=1em},
  boxed title style={size=small, colback=white, colframe=darkgreen, sharp corners},
  top=12pt,
  bottom=6pt,
  left=8pt,
  right=8pt,
  title=#1
}

\title{\textsf{Lessons from the trenches on evaluating machine learning systems in materials science}}

\author[1, $\star$]{Nawaf~Alampara~\orcidlink{0009-0001-7697-7315}}

\author[1, $\star$]{Mara~Schilling-Wilhelmi~\orcidlink{0009-0007-4392-5918}}

\author[1,2,3,4, \Letter]{Kevin~Maik~Jablonka~\orcidlink{0000-0003-4894-4660}}

\affil[1]{Laboratory of Organic and Macromolecular Chemistry (IOMC), Friedrich Schiller University Jena, Humboldtstrasse 10, 07743 Jena, Germany}

\affil[2]{Center for Energy and Environmental Chemistry Jena (CEEC Jena), Friedrich Schiller University Jena, Philosophenweg 7a, 07743 Jena, Germany}
\affil[3]{Helmholtz Institute for Polymers in Energy Applications Jena (HIPOLE Jena), Lessingstrasse 12-14, 07743 Jena, Germany}

\affil[4]{Jena Center for Soft Matter (JCSM), Friedrich Schiller University Jena, Philosophenweg 7, 07743 Jena, Germany}

\affil[$\star$]{These authors contributed equally.}

\affil[\Letter]{\texttt{mail@kjablonka.com}}

\begin{document}
\maketitle

\begin{abstract}
Measurements are fundamental to knowledge creation in science, enabling consistent sharing of findings and serving as the foundation for scientific discovery.
As machine learning systems increasingly transform scientific fields, the question of how to effectively evaluate these systems becomes crucial for ensuring reliable progress.

In this review, we examine the current state and future directions of evaluation frameworks for machine learning in science. We organize the review around a broadly applicable framework for evaluating machine learning systems through the lens of statistical measurement theory, using materials science as our primary context for examples and case studies.
We identify key challenges common across machine learning evaluation such as construct validity, data quality issues, metric design limitations, and benchmark maintenance problems that can lead to phantom progress when evaluation frameworks fail to capture real-world performance needs.

By examining both traditional benchmarks and emerging evaluation approaches, we demonstrate how evaluation choices fundamentally shape not only our measurements but also research priorities and scientific progress.
These findings reveal the critical need for transparency in evaluation design and reporting, leading us to propose evaluation cards as a structured approach to documenting measurement choices and limitations.

Our work highlights the importance of developing a more diverse toolbox of evaluation techniques for machine learning in materials science, while offering insights that can inform evaluation practices in other scientific domains where similar challenges exist.
\end{abstract}

\begin{figure*}
    \includegraphics[width=\textwidth]{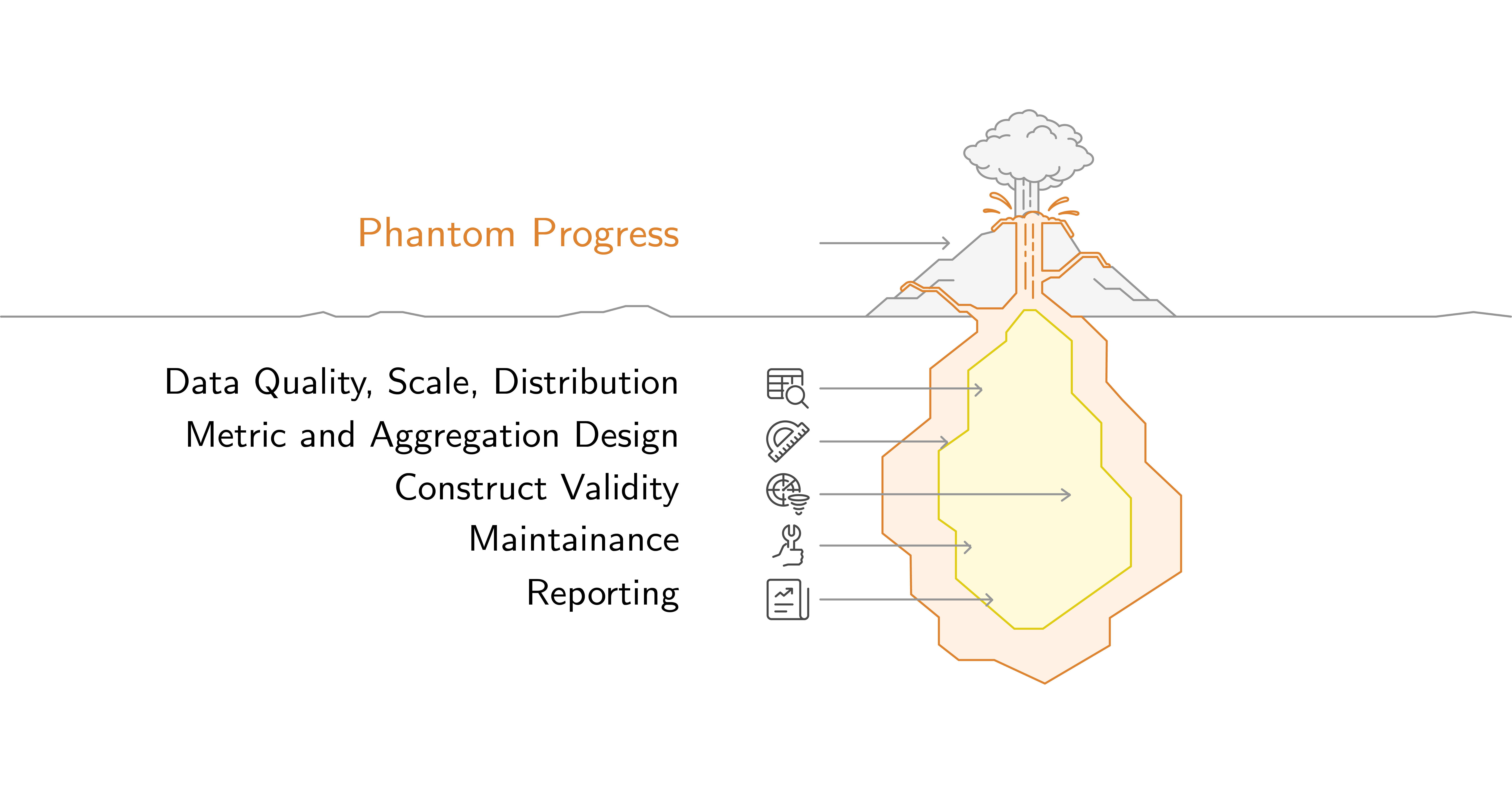} \\ 
    \textbf{The machine learning model evaluation volcano.} Many, if not most, of the factors determining the outcome of the evaluation of a machine learning model are hidden. Many of these hidden design choices lead to \enquote{phantom progress}: While there might be some improvement on a leaderboard, there is no improvement in impact on real-world applications.
\end{figure*}

\clearpage

\section{Introduction}

In science, measurement is fundamental to knowledge creation: \enquote{If you cannot measure, your knowledge is meager and unsatisfactory} (Lord Kelvin). 
Historically, scientific fields focused on developing increasingly refined instrumentation for measuring physical phenomena.
Those measurements not only enabled the sharing of knowledge and collaboration but led to the development of new concepts and scientific discoveries. 
On a larger scale, consistent measurements were foundational for the establishment of markets by creating a shared understanding between buyers and sellers and allowing for consistent pricing mechanisms.\autocite{Hand_2016}

For a measurement to be useful, it must be stable over time and in different environments. 
This can be seen, for example, in the care given to defining the standard units---from carefully watched physical objects, such as Le Grand K, to redefinitions in terms of physical constants to increase stability. 
This stability is not trivial to achieve for learning systems---especially if they can interact with an environment---as the environment could change or the system could learn from the questions it is being probed on.

Such systems that can learn are increasingly being used in materials science.\autocite{butler2018machine, De_Luna_2017}
While many efforts focused on predicting (scalar) properties of crystal structures, new systems are increasingly agentic and take actions in an environment, making their evaluation surprisingly difficult. 
In addition, as the community strives to address grand challenges, our interests shift more and more to the performance of a material in its application in the real world (in contrast to predicting the properties of an idealized crystal structure).\autocite{Moosavi_2020_role} \\

\begin{figure}
    \centering
    \includegraphics[width=1\linewidth]{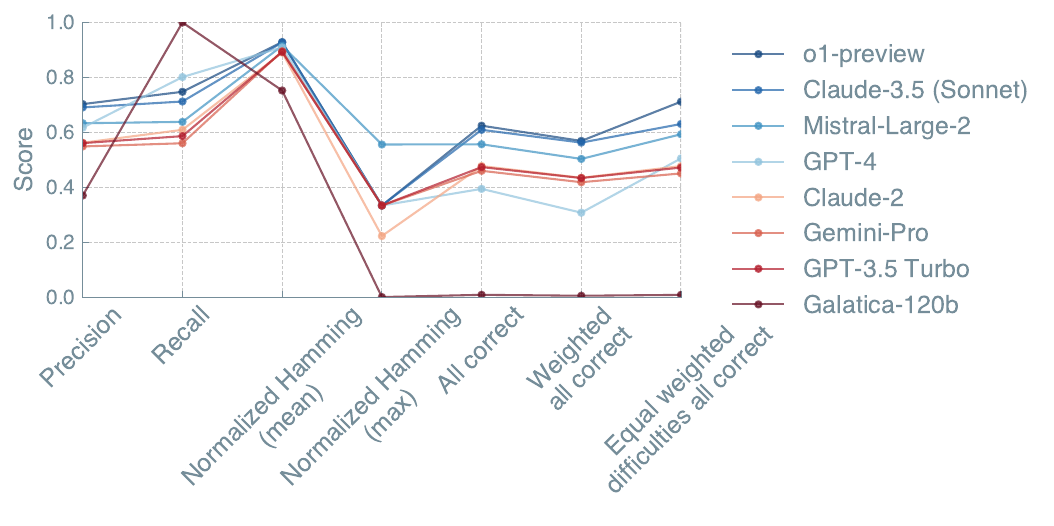}
    \caption{\textbf{ChemBench \autocite{mirza2024largelanguagemodelssuperhuman} ranking based on different scoring metrics.} All metrics are a sum, weighted sum, or maximum values over all multiple-choice questions. The weighted sums are calculated by taking the manually rated difficulty (basic, immediate, advanced) of the question into account. For equal weighting, all categories are weighted even, regardless of the number of questions. The metric all correct is a binary metric indicating if a given answer is completely correct. For normalized Hamming (max), the normalized maximum value of the Hamming loss of each model was taken. We find that the ranking of models changes if we change the metric, or even just the aggregation --- showcasing the importance of proper and transparent design of evaluation suites.}
    \label{fig:ranking_changes}
\end{figure}

\noindent It is important to note that the act of measurement and the act of making a choice are inseparable. 
While a system could theoretically be measured in infinite ways, any evaluation framework ultimately emerges from decisions about what to measure and how to measure it. 
These decisions --- whether made by individual researchers or evaluation framework developers --- make any evaluation inherently a social construct, shaped by human choices and values.\autocite{Dotan_2020} 
Often, these choices and values are implicit and not communicated in a transparent and systematic form --- which can impede progress. \Cref{fig:ranking_changes} gives such examples of how the choice of metric or aggregation of metrics can fundamentally change the ranking of models and conclusions one might draw from a model evaluation.

\begin{figure}
    \centering
    \includegraphics[width=1\linewidth]{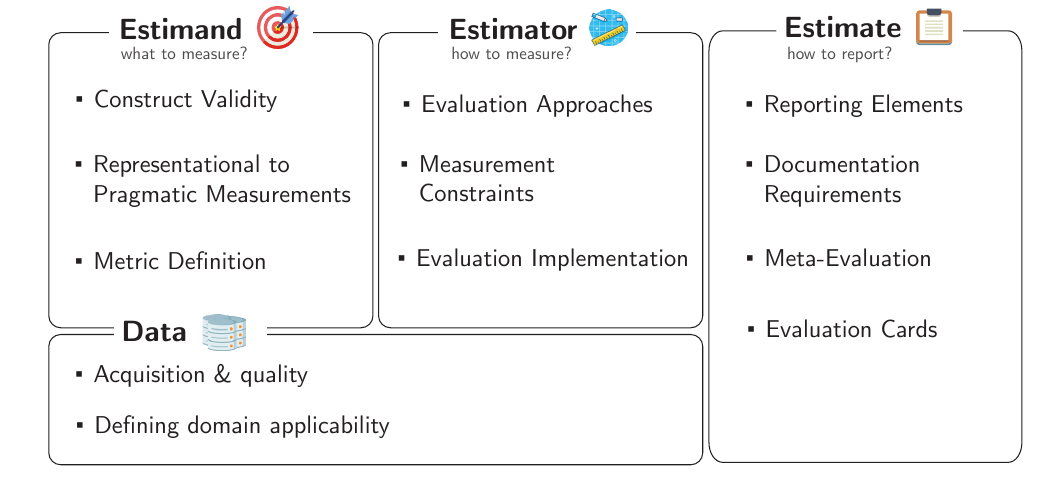}
    \caption{\textbf{Estimand, estimators, estimate:} The diagram illustrates a conceptual framework for machine learning system evaluation in materials science, structured around three key components outlined in this article. Often, both estimand and estimator depend on data. This highlights how the boundaries between what we measure and how we measure it are not always clearly delineated in the evaluation of complex machine learning systems for materials science. }
    \label{fig:overview}
\end{figure}

Given the importance of measurements and the many pitfalls surrounding it, this article reviews the science of evaluating machine learning systems in materials science through the lens of statistical measurement theory. 
We organize our discussion around three fundamental concepts (see \Cref{fig:overview}): estimands (what we want to measure), estimators (how we measure it), and estimates (the measurement results). 
The discussion attempts to not overly focus on one type of model evaluation (e.g., benchmarks) but to highlight overarching themes.
We hope this overview helps developers and users of evaluation frameworks to avoid our mistakes and perhaps even sparks a \enquote{science of materials science evals.}

\section{Defining Estimands: What Are We Measuring?}

\subsection{Construct Validity}

One of the most fundamental challenges in designing measurements is ensuring construct validity --- that we measure what matters for real-world applications. 
Many measurements, especially benchmarks, suffer from what is known as the McNamara Fallacy:\autocite{Basler_2009}

\begin{enumerate}
    \item Measure what can be easily measured
    \item Disregard that which cannot be measured easily
    \item Presume that which cannot be measured easily isn't important
    \item Conclude that which can't be easily measured doesn't exist
\end{enumerate}

That is, measurements sometimes tend not to be designed based on what evaluations would matter most --- but based on what evaluations are most easy to construct. 
This can lead to \enquote{phantom progress,} where metrics improve on benchmarks without corresponding improvements in real-world performance.\autocite{dacrema2019really} 
Falling for this fallacy also indirectly boosts models which do well on data that is easy to collect on scale.\autocite{Dotan_2020} 

\begin{examplebox}[Examples for the McNamara Fallacy]
\begin{itemize}
    \item \emph{Overfocus on \gls{mcq}} Many benchmarks are only \gls{mcq}, since they are easy to score.
    \item \emph{Overfocus on Prediction Accuracy:} Many ML models in materials science are evaluated primarily on metrics like \gls{rmse} or $R^2$ for property prediction, while ignoring interpretability, physical consistency, or domain applicability. A model might achieve excellent accuracy on test data but fail to capture fundamental physics or provide actionable insights for materials design.
    \item \emph{Ignoring Synthesis Feasibility:}  Models for materials discovery are often benchmarked solely on property prediction accuracy without considering whether predicted materials can actually be synthesized. A \enquote{high-performing} model might predict theoretically excellent but practically impossible materials.
    \item \emph{Formation Energy vs. Synthesizability:} Similarily, some \gls{ml} models for materials discovery are evaluated solely on formation energy prediction accuracy, ignoring critical aspects like kinetic stability, precursor availability, and reaction pathways that determine whether materials can actually be synthesized.
    \item \emph{Dataset Size Over Quality:} Properties for which large datasets exist (e.g., computational bandgaps) tend to receive more attention than properties for which only smaller ones exist (e.g., experimentally determined stereoselectivity). 
\end{itemize}
\end{examplebox}

\begin{figure}
    \centering
    \includegraphics[width=1\linewidth]{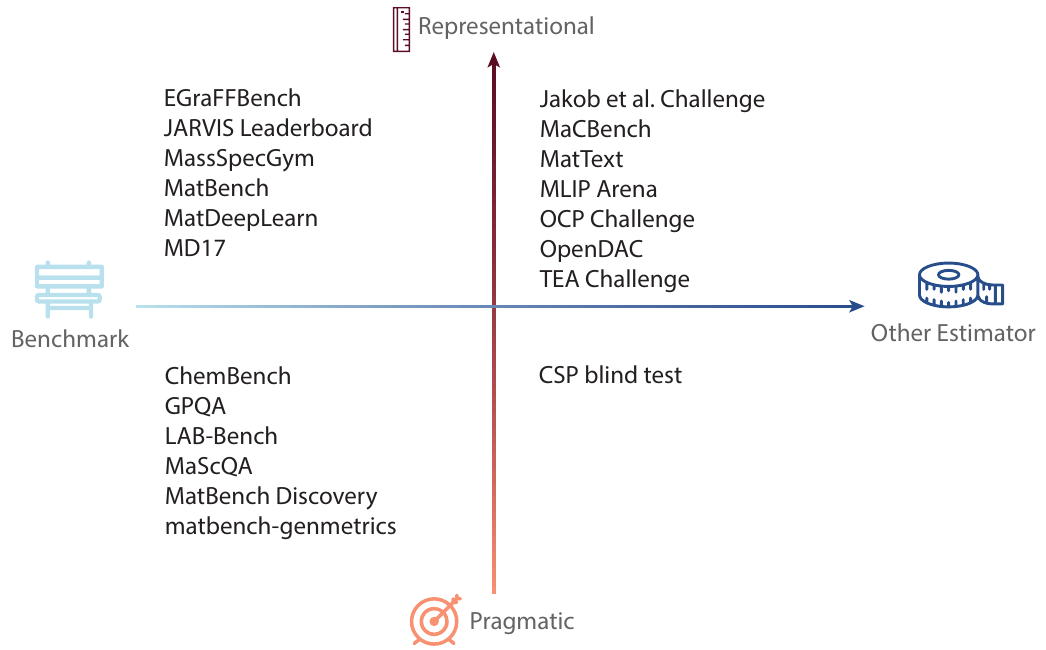}
    \caption{\textbf{Common materials science benchmarks and other estimators on the continuum from representational to pragmatic:} To visualize this conceptual distinction, we heuristically positioned selected benchmarks along two axes: the horizontal axis reflects how representational (i.e., measuring intrinsic, task-independent properties) versus pragmatic (i.e., task-constructed, decision-driven measures) the evaluation is intended to be; the vertical axis distinguishes benchmarks from other estimators or evaluation tools. 
The classification is based on qualitative assessment by the authors and should be interpreted as illustrative rather than definitive. For simplicity, we group the evaluation frameworks in quadrants. In reality, however, they live on a continuum. Representational Benchmarks: Representational benchmarks include EGraFFBench \autocite{https://doi.org/10.48550/arxiv.2310.02428}, JARVIS Leaderboard \autocite{https://doi.org/10.5281/zenodo.14212326,  Choudhary_2024}, MassSpecGym \autocite{https://doi.org/10.48550/arxiv.2410.23326}, MatBench \autocite{Dunn_2020}, MatDeepLearn \autocite{MatDeepLearn}, and MD17 \autocite{Chmiela2017}. On the pragmatic end of the spectrum, relevant benchmarks include ChemBench \autocite{mirza2024largelanguagemodelssuperhuman}, GPQA \autocite{https://doi.org/10.48550/arxiv.2311.12022}, LAB-Bench \autocite{https://doi.org/10.48550/arxiv.2407.10362}, MaScQA \autocite{https://doi.org/10.48550/arxiv.2308.09115}, MatBench Discovery \autocite{riebesell2024matbenchdiscoveryframework}, and matbench-genmetrics \autocite{Baird2024}. Other representational estimators include the Jakob et al. Challenge \autocite{Jakob2025}, MaCBench \autocite{alampara2024probinglimitationsmultimodallanguage}, MatText \autocite{https://doi.org/10.48550/arxiv.2406.17295}, MLIP Arena \autocite{Chiang_MLIP_Arena}, OCP Challenge \autocite{Chanussot_2021}, OpenDAC \autocite{Sriram_2024}, and the TEA Challenge \autocite{Poltavsky_2025}. On the pragmatic side, an additional estimator is the CSP blind test \autocite{Lommerse2000}.}
    \label{fig:representational_to_pragmatic}
\end{figure}

\subsection{From Representational to Pragmatic Measurements} 
To more systematically describe the design of measurements, it can be helpful to think of a continuum of measurements from \enquote{representational} to \enquote{pragmatic} (\Cref{fig:representational_to_pragmatic}). 
Representational measurements aim to capture inherent properties that exist independently of measurement (like length or mass). 
In contrast, pragmatic measurements are explicitly constructed to serve practical purposes and aid decision-making (like credit scores or school grades) but not to measure properties that exist independent of the measurement.\autocite{Hand_2004} 
That is, pragmatic measurements are defined through the collection of measurements themselves. 
This has important implications: Unlike representational measurements that can be validated against physical reality, pragmatic measurements can only validate themselves through their utility in providing meaningful insights.

Once established, pragmatic measurements can influence the system they aim to measure. 
For example, academic metrics like the h-index don't just measure scholarly impact --- they shape how researchers publish and what they consider impactful work (Goodhart's law). 
The measurement becomes part of the phenomenon it measures.
Similar behavior can be observed for machine learning benchmarks.\autocite{hsia2023goodharts, teney2020value} 
This phenomenon is linked to the fact that measurements are dimensionality reductions: Due to the dimensionality reduction, one can optimize the final scores in ways that do not optimize the underlying quality the measurement was supposed to measure. 
This phenomenon is also known as reward hacking.\autocite{Bengio_2024} 

In addition, because pragmatic measurements get their meaning from the corpus of evaluation tasks, they are inherently comparative. For example, a student's test score is meaningful only in relation to other scores in the database, not as an absolute measure of knowledge.

That is, as one moves from representational to pragmatic measurements, there are caveats in the design and interpretation of the measurement one has to keep in mind. 
This is something we observed in our own work: We built ChemBench --- a benchmark designed for evaluating the chemical capabilities of \glspl{llm} --- in the first iteration by collecting diverse questions from various sources, such as university exams.\autocite{mirza2024largelanguagemodelssuperhuman} 
However, this relatively unstructured data-collecting process makes it difficult to pinpoint the skills that this benchmark measures. 
As with most other \gls{llm} benchmarks, one can use ChemBench to compare different models --- but this does not necessarily connect to deeper properties or qualities of these models. 
In a follow-up work targeted at evaluating multimodal models (MacBench) we attempted to improve on this and constructed tasks to answer specific hypotheses --- pushing the benchmark to be more representational.\autocite{alampara2024probinglimitationsmultimodallanguage}
In this way, we have a way more specific lens on an aspect of a system we study---which can lead to more actionable insights.

In addition, such a more representational design can also help avoid problems such as \gls{harking}. 
This practice, where a hypothesis is formulated after analyzing the results of a study but is presented as if it were conceived beforehand (a priori), skews the scientific record. 
It increases the likelihood of false positives being published, as exploratory findings are elevated to the status of confirmatory evidence without proper validation.\autocite{gencoglu2019harkdeeplearning, Kerr_1998, Varoquaux_2022} 
Practically, this can be thought of as $p$-hacking: If we just collect the data that is easiest to collect for constructing our evaluations and then try to find some patterns in there, we will find some pattern at some point just by chance.\autocite{gundersen2022sources}

\begin{examplebox}[Example for HARKing]
HARKing often manifests when researchers run multiple model variants against various metrics, then retrospectively claim they were specifically targeting the benchmark where they achieved the best results. For instance, a team might train models with different hyper parameters on multiple datasets (e.g., MatBench, OCP, OpenDAC), find their approach works particularly well on OCP, and then frame their paper as if improving OCP performance was their primary objective all along. This selective reporting creates a misleading impression of intentional progress rather than chance discovery.  Similarly, researchers might test against multiple metrics (e.g., accuracy, F$_1$, AUC-ROC) but only highlight those where their model excels, presenting discoveries as predetermined research goals.
\end{examplebox}

\subsection{Data Acquisition and Quality}

The continuum from representational to pragmatic measurements becomes relevant when considering the role of data in
evaluations of \gls{ml} systems. 
In most evaluations, datasets have a dual role: both defining what we measure (e.g. via the distribution of the test set) and providing the mechanism for measurement (due to the fact that we test on a test set).
This is different from representational measurements, where the property being measured exists independently of the measurement apparatus.

This dual role can lead to problems in the interpretation of evaluation results in materials science, where, for example, benchmark datasets are created at different levels of theory (e.g., \gls{dft}, force fields, experimental measurements) and compiled from different sources (e.g., literature, high-throughput calculations, experimental databases). 
Each data source not only brings its noise and biases but subtly redefines what capability we are measuring. 
For instance, a model that performs well on \gls{dft}-derived data might be learning something quite different from one that excels on experimental measurements, even if both claim to predict the same property.

\begin{examplebox}[Example: Data Defines What We Measure]
\begin{itemize}
    \item \emph{Bandgaps:} Models trained on \gls{dft}-calculated bandgaps (typically using \gls{pbe} functionals) learn to predict values that systematically underestimate experimental bandgaps.\autocite{Lee_2016} 
A model achieving \enquote{high accuracy} on \gls{dft} bandgap predictions is actually learning this systematic underestimation pattern rather than true experimental bandgap values. Conversely, models trained on experimental bandgap measurements learn different patterns related to measurement conditions and material synthesis variations. When researchers compare these models using the same evaluation metrics (\gls{mae}, \gls{rmse}), they create a false equivalence --- the \enquote{\gls{dft}-accurate} model might show excellent metrics on \gls{dft} test sets while being consistently off on real materials.\autocite{Marzari_2016, Jain_2016, Maurer_2019, Jha_2019} 
    \item  \emph{Catalysis:} In catalysis, models trained on clean CO adsorption energies from DFT often fail when applied to noisy, high-throughput experimental data due to surface reconstruction or coverage effects that are absent from the training domain. \autocite{Ulissi2017}
\end{itemize}
\end{examplebox}

\paragraph{Sampling} 
In an ideal world, we would like to know how our model performs in all possible cases, for example, on all possible materials. 
Testing on all imaginable materials, however, is clearly not feasible. 
Hence, one needs to perform a selection of data on which we measure the performance of our \gls{ml} system.
Since this dataset defines what we measure, it is important to have a well-defined and systematic protocol for constructing it.

This requires thought on how representative the data is of the real population we care about.\autocite{Kpanou_2021, Sheridan_2013, riebesell2024matbenchdiscoveryframework} 
At the same time, we also need to keep in mind that we need to have enough data to have sufficient statistical power.\autocite{Schmidt_2024}
Balancing those desired properties with real-world constraints leads to biases.
For example, databases, by construction or computational constraints, might be limited to smaller structures (e.g., the initial version of QMOF\autocite{Rosen_2021, Rosen_2022} focused on smaller unit cells) or to structures that pass certain validation frameworks.\autocite{Chung_2019, Zhao_2024} 
Such biases\autocite{Jia_2019,  Fujinuma_2022} are well described for data in patents, where organic reactions tend to be biased to a small number of reaction types, such as amide formations or couplings. 
Another prominent bias is that failed experiments are seldom reported in the literature.\autocite{Schneider_2016, Brown_2015, griffiths2021dataset, Raccuglia_2016} 
A good performance of a model on those datasets might thus not translate to good performance on many practical use cases in the \enquote{real world.}\autocite{bartel2020critical, fujinuma2021reflections}

\begin{examplebox}[Example: Bias in Test Data]
\begin{itemize}
    \item \emph{Generative Modeling:} MP-20\autocite{xie2021crystal, Jain_2013} is commonly used in generative modelling of crystal structure.  This dataset predominantly contains simple structures with unit cells including at most 20 atoms and limited chemical diversity. As a result, models might show impressive performance on this dataset but fail entirely when tasked with generating complex metal-organic frameworks or structures containing rare earth elements. The bias toward computationally manageable structures creates an artificial environment where models can achieve high metrics (stable structures scores) while being fundamentally incapable of generating the diverse, complex structures needed for many real-world applications.
    \item \emph{Disorder Modeling Limitations:} AI-driven materials discovery systems tend systematically overpredict ordered structures because \gls{dft} calculations at \SI{0}{\kelvin} don't account for configurational entropy, leading to benchmark evaluations that don't reflect experimentally achievable outcomes.\autocite{Cheetham_2024, Leeman_2024}
\end{itemize}

\end{examplebox}

\begin{examplebox}[Example: Overconfidence by Sampling Overly Broad Distributions]
In some cases, datasets span a very wide range of properties. For example, the ESOL\autocite{Delaney_2004} dataset in MoleculeNet\autocite{wu2018moleculenet} spans more than 13 logs. With this wide property range one can obtain (seemingly) good correlations --- even though those models might fail to resolve differences on the more meaningful smaller range of properties observed in drug discovery.\autocite{walters2023benchmarks}
\end{examplebox}

\paragraph{Diversity and Duplicates}

Linked to this are diversity\autocite{Moosavi_2020, Majumdar_2021} and duplication. 
For most evaluations, one should aim for a diverse set of tasks (e.g. structures for which one aims to predict properties) with independent samples as considerable similarities between the examples in a test set will limit the information one can extract from it. 
Concretely, a model achieving high accuracy in predicting band gaps for 100 nearly identical cubic oxide perovskites (with correlation coefficient $\rho  \approx 0.99$ between samples) effectively provides only in the order of $\sqrt{1-\rho^2} \cdot 100 \approx 14$ independent test points.  While such a model may be valuable for users focused solely on perovskites, its benchmark performance says little about generalization to other material classes---such as layered chalcogenides---unless tested accordingly. This highlights the need for benchmark datasets that reflect the intended scope of application.

Ensuring this is not trivial, as the definition of what makes two systems equivalent or identical depends on the problem (i.e., which degrees of freedom can be ignored). 
In some cases, only composition information is needed to describe a particular phenomenon.\autocite{tian2022information}
Therefore, defining duplication based on the exact overlap of atomic positions would be a too strict definition of a duplicate. As a consequence different heuristics have been proposed to identify duplicated materials, such as graph hashes on labeled, unlabeled, or even just scaffolds of structure graphs.\autocite{Jablonka_2023, Barthel_2018, Hart_2024, siron2025lematbulk, Ongari_2022}

Identifying what data is duplicated or what a corpus entails is even more difficult in the case of \glspl{llm}, where the cost of checking for duplication is high due to the size of training corpi as well as the fact that one needs to go beyond checks for exact string matches to semantic matches (in potentially different languages).\autocite{brown2020language} 
Yet, even if one were able to perform these computations, it is often not feasible to do so as the training datasets for the best performing models are frequently not disclosed.\autocite{biderman2024lessonstrenchesreproducibleevaluation, balloccu2024leak}

\paragraph{Automatically Generated Data} 
A growing body of work uses \glspl{llm} to generate questions (and sometimes even answers).\autocite{song-etal-2023-honeybee, fang2023mol, xie2024darwin} This raises an interesting philosophical point: Can we even measure performance that is stronger than the one of the models we used for generating the questions? 
The answer to this question is unclear, and currently hybrid systems perform best \autocite{shah2024aiassisted}. 
Thus, skepticism for evaluations that are completely based on automatically generated data is still well placed.
However, the inverse question \enquote{how do we test models that are better than the best humans} is being actively researched, too, and is an important topic for AI alignment (known as scalable oversight).\autocite{bowman2022measuring, amodei2016concrete} 
Given some parts of the chemical sciences frontier models already outperform experts\autocite{mirza2024largelanguagemodelssuperhuman} this question will also become increasingly relevant for materials science and chemistry.

\paragraph{Stability, Uncontrollable Variables, Invariances}

For some tasks, e.g., agent evaluations, the environment they interact with is not stable over time. 
In a simple case, there might be rate limits for some websites, and then the order in which tasks are executed matters (e.g., a first task might already lead to blocked requests, which causes a second task for a materials science data mining agent to fail). \autocite{kapoor2024aiagentsmatter} 
This is a specific example of the fact that environments in which we or an agent operate might change --- either because of the agent’s actions or some other factors. 
For this reason, measurements of agentic systems in complex environments are not necessarily invariant to time, space, and execution order. 
In practice, this increases the need for transparency on how measurements have been performed. 

\begin{examplebox}[Examples: Uncontrolled Variables in Machine Learning]
\begin{itemize}
    \item  \emph{Material Degradation:} When evaluating perovskite solar cell efficiency, measurements taken immediately after fabrication versus after several hours of operation yield different results due to inherent material degradation. This time-dependent performance can make it difficult to compare approaches (e.g., active learning experiments, or agentic experimentation).
    \item \emph{Developing Databases:} A benchmark for crystal structure prediction might rely on the Materials Project database,\autocite{Jain_2013} which undergoes periodic updates of structures, properties, and calculation methods. \gls{ml} models evaluated against version 2022.10 might show different performance compared to those tested on version 2023.05, making temporal comparison of published results problematic without explicit version tracking.
\end{itemize}
\end{examplebox}

Transparency is particularly relevant when proprietary, closed source,  frontier models (such as OpenAI's \glspl{llm}) are used via an \glspl{api}.\autocite{Palmer_2023, closed_ai_baselines, Spirling_2023} 
The exact systems that are deployed behind such an \gls{api} might change at any given time\autocite{chen2023chatgpt2s} and might also be deprecated by the provider. 
Thus, when proprietary models are used, the interaction needs to be stored along the timestep when the experiment has been performed.
It is also important to realize that the use of such closed models as baseline is often not suitable---given that we know little about architecture and training details, we cannot use them as a way to measure methodological improvements.\autocite{closed_ai_baselines}
On top of that, in certain usage mode, the data we send via an \gls{api} or web interface might be used for training the model --- making the measurement we aim to perform unstable and incomparable to other systems.

\paragraph{Data Problems Might not be Fixable --- But Require Transparency}

Many of the issues described above, such as data scale and diversity, are hard to \enquote{fix.} 
Many problems compete on a Pareto frontier: Within a fixed budget, data fidelity and data scale are often competing objectives. We need to make tradeoffs. 
However, even when we might not have solutions for these problems, we can --- and must --- be transparent about them if we aim to follow the scientific method.
Thus, designers of model evaluation suites could increase the impact and value of their developments by providing comprehensive discussions of limitations and tradeoffs they made in the design of the measurement instrument.

\subsection{Metric Definition}

\paragraph{Defining a Metric is a Dimensionality Reduction}

The definition of metrics requires careful consideration. 
If we take the example of classification, we could, in principle, (manually) analyze every single prediction on our test examples. 
This, however, is, in most cases, impossible to do and to communicate. 
Thus, we define metrics that \enquote{summarize} the performance in one or a few numbers. 
The choice of these numbers defines what we measure. 
While the \gls{mae} or classification accuracy are common choices, they might not always align with our scientific or application goals \autocite{blagec2020critical, Opitz_2024}. 
For materials discovery applications, for instance, we might care more about correctly identifying promising candidates in the top percentile rather than accurate predictions across the entire chemical space \autocite{Jablonka_2023}. 
Thus, metrics like discovery yield or discovery probability have been proposed \autocite{Borg_2023}.
This is a specific example of a general fact: Different errors have different costs associated with them if we were to apply the \gls{ml} system in the real world. 
Most commonly, however, we do not account for this in the design of metrics. 
In addition, different metrics have different statistical properties. 
For instance, one might desire that a metric shows monotonicity: That is, that a correct prediction does not make the score worse and that an incorrect prediction does not make it better: A property that not all widely used metrics fulfill. 
One might also hope that a metric allows for distinguishing random guessing from informed decision-making. Again, this is not a property that all widely used metrics fulfill.\autocite{DBLP:journals/tacl/Opitz24} \\

\begin{examplebox}[Example: Suboptimal Metrics in Materials Science]
\begin{itemize}
    \item \emph{Validity Metric in Crystal Structure Generation:} The \enquote{validity} metric in crystal structure generations exemplifies an oversimplified evaluation --- considering a structure valid if atoms are merely \SI{0.5}{\angstrom} apart\autocite{Court_2020} sets such a low bar that models can achieve high scores without generating physically realistic structures. This metrics are saturated with the current best performing models and should be deprecated.
    \item \emph{Materials Discovery Based on Goodness of Fit:} Relying solely goodness of fits of Automated Rietveld analyses led to claims of discovery that were actually known materials. This is because a proper analysis would need to prove that the fit is better than the one of any other phase that might have been formed. Even less ambigous might be the use of \gls{scxrd} instead of \gls{pxrd}.\autocite{Leeman_2024} 
\end{itemize}

\end{examplebox}

\noindent Designing good metrics is complex in materials science, where we deal with multi-objective optimization problems, as a material's performance rarely being determined by a single property.\autocite{Jablonka_2021, DBLP:conf/icml/Wagstaff12} 
In practice, designing materials for real world applications even requires an understanding of techno-economic and engineering requirements (scalability, cost, and application-specific constraints) to cross the \enquote{valley of death} separating materials discovery from commercialization.\autocite{Moosavi_2020_role, Charalambous_2024, Markham_2002}

Being aware of this is essential because some shortcut metrics (e.g., materials properties) might not correlate at all with the final (perceived) performance in a real-world application (e.g., incorrectly using formation energy as a proxy for stability).\autocite{Burns_2020, Bartel_2020, Sun_2016} 
From this point of view, an increasing number of works attempt a more holistic assessment of materials discovery.\autocite{Charalambous_2024, Burns_2020}

\paragraph{Dealing with the Fact That We Cannot Always Specify What Matters}
However, from a conceptual point of view, also these works struggle with the fact that there is no reliable and general framework to deal with the fact that for novel applications or scientific discoveries, we often do not even know what good metrics or (\enquote{stepping stones}) would be before we fully understand the application.\autocite{stanley2015greatness, wang2019pairedopenendedtrailblazerpoet, secretan2008picbreeder, Schrier_2023} 
In broad strokes, one could say that before having the discovery, we do not know what could bring us there. 


\paragraph{Aggregations} 
A subtle but important point is that we typically have many different tasks in one evaluation (i.e., scores for different subsets of problems). 
Or we might have multiple classes for a given classification problem. 
In any case, we want to summarize the metrics we computed even further. 
Again, this is part of the definition of what we measure and thus can impact the overall ranking of models.\autocite{binette2024improvingvaliditypracticalusefulness, dehghani2021benchmarklottery}

While this effect is commonly discussed in the context of multiclass classification metrics --- for example, showcasing how micro averaging and macro averaging emphasize different aspects (e.g., minority classes) --- this is less discussed when aggregating benchmark results. 
One phenomenon one might be confronted with is that averaging over tasks with different performance ceilings might lead to the artificial overweighting of tasks with a larger range.
This becomes relevant in materials science when certain tasks have inherent performance ceilings due to data quality issues.

\begin{examplebox}[Example: Impact of Aggregations]
Consider two different machine learning tasks. The first is a noisy task where even human experts can only achieve 40\% accuracy due to inherent limitations like experimental uncertainty or labeling noise. The second is a clean task where near-perfect performance is possible. If we simply average performance across these tasks, we create a misleading picture. On the noisy task, a model improving from 35\% to 38\% accuracy has closed 15\% of the remaining gap to human performance (3\% improvement out of a possible 5\%). On the clean task, a model improving from 90\% to 95\% accuracy appears numerically larger (5\% improvement), but represents a smaller relative advancement in terms of closing the gap to optimal performance. Simple averaging would give more weight to the clean task, overlooking the potentially more significant breakthrough on the harder, noise-limited task. This illustrates why evaluation methods need to account for the different characteristics and difficulty levels of various tasks, rather than treating all percentage improvements equally.
\end{examplebox}

\noindent For aggregations, too, there is not one correct choice, and being transparent about the choices and the reasons for those choices is the best recommendation one can make. 
To increase transparency and make analyses more systematic, recent works have investigated the use of test theory, such as item response theory, to more systematically combine scores on different subsets of an evaluation suite.\autocite{mattos2021assessment, van2016handbook, schilling-wilhelmi2025lifting}

\subsubsection{Constraints}

The constraints placed on model evaluation frameworks play a crucial role in defining what is being measured. 
These constraints shape not only how models can be developed and tested but also influence what capabilities we can meaningfully compare. 

\paragraph{Training Data Constraints} 

Different evaluation frameworks take varying approaches to restricting training data usage. Some, like the Open Catalyst Project (OCP)\autocite{Chanussot_2021} and MatBench,\autocite{Dunn_2020} impose strict limitations on allowed training data sources to ensure fair comparisons. 
Others, like MatBench Discovery,\autocite{riebesell2024matbenchdiscoveryframework} adopt a more permissive stance to encourage innovative approaches.
These choices fundamentally impact what we measure: strict data constraints help isolate specific modeling capabilities, while looser constraints might better reflect real-world deployment scenarios where models can leverage diverse data sources.

\paragraph{Computational Resource Constraints }

Given that for some modeling paradigms, such as \glspl{llm}, performance can be improved by a large margin by spending more compute power --- not only in training but also in test time --- there is a need for reporting or even controlling the use of computational resources.
Only transparency about the computational resources that have been used for solving a given task can allow us to distinguish a brute-force solution from an efficient data-driven one. 
Since many, if not most, currently used systems (including agents) are stochastic, simply averaging over multiple runs will trivially increase performance. 
For instance, the performance of AlphaCode increases from close to 0\% zero-shot to over 15\% with 1,000 retries and over 30\% with a million retries.\autocite{Li_2022} 

Notably, computational constraints could be imposed both in training and inference time. 
For instance, MatBench Discovery cites the wish for scalable solutions as one of the motivations for using a large test set.\autocite{riebesell2024matbenchdiscoveryframework} 

Computational resource budgets are directly related to hyperparameter optimization. 
The degree of allowed hyperparameter optimization represents a critical constraint that directly impacts measured performance. 
Without clear constraints on hyperparameter optimization budget, performance comparisons can become dominated by differences in optimization resources rather than fundamental model capabilities. 
Often, it has been found that researchers spend more effort in optimizing the hyperparameters for \enquote{their} approach than for baselines.\autocite{liao2021are} 

Importantly, the use of computational resources also links to broader impact of AI in terms of energy use and centralization of power. \autocite{Luccioni_2024, strubell-etal-2019-energy}

\paragraph{Reproducibility Constraints}

The designers of an evaluation suite can choose to request a certain level of reproducibility, e.g., via the sharing of code that can be used to reproduce the results. For example, MatBench discovery,\autocite{riebesell2024matbenchdiscoveryframework} Matbench,\autocite{Dunn_2020} and the JARVIS leaderboard\autocite{Choudhary_2024} require the submission of code that was used to generate the prediction. 
They also have some level of manual oversight as additions to the leaderboard are handled via GitHub Pull Requests that are reviewed by the maintainers. 
While this can ensure a higher level of reproducibility, it raises the bar for both contributors and maintainers. It also limits the degree to which researchers with proprietary code can participate. 
Many evaluation frameworks in the \gls{ml} community thus do not have such requirements and also have a fully automated process for addition to the leaderboard. 
Aiming for the largest possible number of contributions, we also chose this model for ChemBench and MaCBench by building leaderboards on the HuggingFace platform.

\paragraph{Safety Constraints}

For autonomous systems and agents, safety constraints can help ensure responsible development and deployment. 
This might be implemented by defining allowed interaction patterns for systems that might control experimental equipment or make decisions about material synthesis. 
At this point, no standards for such safety constraints exist.\autocite{Wang_2025}

\begin{practicalbox}[Practical Implications: Defining Estimands]
\begin{itemize}
\item Explicitly document what capability you're measuring and how it relates to real-world materials science applications
\item Position your measurement on the representational-pragmatic continuum and adjust validation approach accordingly:
  \begin{itemize}
  \item Representational: Benchmark against physics-based standards
  \item Pragmatic: Define clear criteria relating to practice
  \end{itemize}
\item Assess and document your data's:
  \begin{itemize}
  \item Chemical/structural diversity 
  \item Potential biases in composition, structure, and properties
  \item Independence between training and test examples
  \item Labeling noise and systematic errors
  \end{itemize}
\item Select metrics based on application needs, not convention:
  \begin{itemize}
  \item For discovery: Include ranking and top-N performance
  \item For property prediction: Consider error impact on decisions
  \item For all applications: Use multiple complementary metrics
  \end{itemize}
\item Set clear constraints on what data, resources, and approaches are permitted, documenting how these choices shape what you can meaningfully measure.
\end{itemize}
\end{practicalbox}

\section{Estimator: How Do We Measure?}

\begin{figure}
    \centering
    \includegraphics[width=1\linewidth]{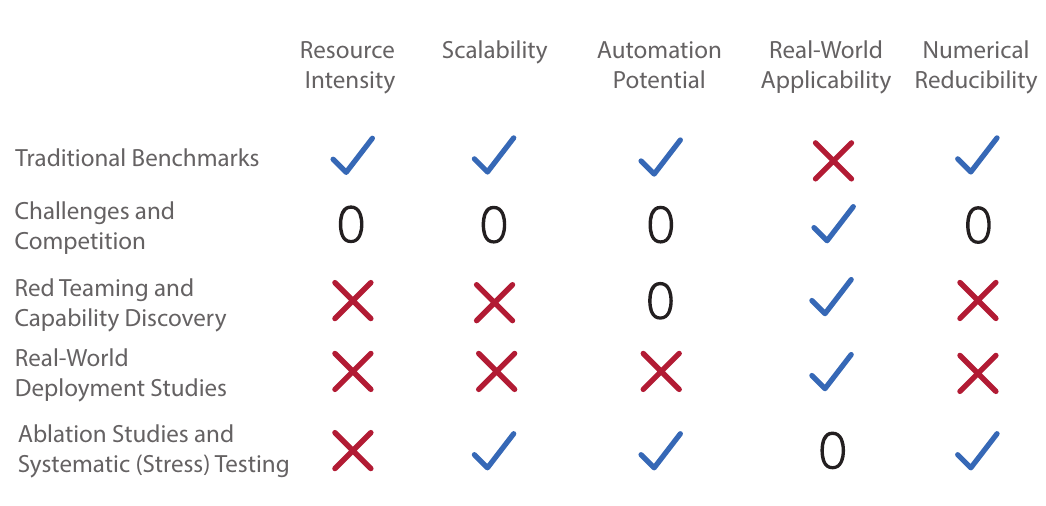}
    \caption{\textbf{Different approaches for evaluating ML systems.}  This figure compares five evaluation approaches across five key dimensions. Traditional benchmarks offer often low resource intensity, good scalability, automation potential, and numerical reducibility, but often lack real-world applicability. Challenges and competitions excel mainly in real-world applicability. Red teaming and capability discovery score moderately in automation potential and strongly in real-world applicability. Real-world deployment studies provide the highest real-world applicability but perform poorly on other dimensions. Ablation studies and systematic testing show strengths in scalability, automation potential, and numerical reducibility but are typically resource intense and have neutral real-world applicability. }
    \label{fig:estimator-types}
\end{figure}

While benchmarks have become the de facto standard for evaluating ML models in materials science, they represent just one approach among many possible evaluation strategies (see \Cref{fig:estimator-types}). 
The choice of evaluation approach fundamentally shapes not only how we measure performance but often also what aspects of model capability we can capture.

\paragraph{Traditional Benchmarks}

The most common approach in materials \gls{ml} involves evaluating models on curated test sets. 
While this enables standardized comparison between models, benchmarks can become outdated (\enquote{saturated}) as the field progresses and may not capture the full complexity of real-world applications.\autocite{Ott_2022} 
Moreover, as discussed before, benchmark datasets play a dual role in both defining and measuring capabilities.
Benchmarks have led to much heralded success in machine learning, but also have been blamed for concerning trends: They can lead to overfitting to the benchmark as well as a too narrow focus on optimizing a small number of quantities the benchmark measures.\autocite{dehghani2021benchmarklottery}

\begin{examplebox}[Example: Benchmark Saturation in Materials Science]
Several materials science benchmarks have reached or are approaching saturation:

\begin{itemize}
    \item \emph{QM9 Dataset:} Best performing models  approach the inherent noise level of the \gls{dft} calculations used to generate the dataset. Further \enquote{improvements} often reflect overfitting rather than genuine advances.
    \item \emph{MatBench Discovery:} Despite being newer, the best models are now $F_1>0.9$, making it difficult to meaningfully distinguish between top-performing approaches.
\end{itemize}
\end{examplebox}

\paragraph{Challenges and Competitions}

Time-bounded competitions, such as \gls{casp},\autocite{moult2005decade} offer a different evaluation paradigm.
These events commonly focus on prospective prediction rather than retrospective evaluation. 
For this reason, they also limit the risk of overfitting. 
They typically also focus on evaluating performance in a way that is more closely related to real-world deployment.
This can provide deeper insights into model capabilities but requires significant community coordination and a trusted third party managing the coordination.  

In materials science, such a challenge has, for instance, been hosted by the Novel Materials Discovery (NOMAD) Centre of Excellence\autocite{ghiringhelli2017towards} on the Kaggle platform with the goal of predicting properties of \glspl{tco}.\autocite{Sutton_2019}


\paragraph{Red Teaming and Capability Discovery}

As the \gls{ml} systems we build become more complex, \autocite{M_Bran_2024, Boiko_2023, darvish2024organa}  it becomes increasingly difficult to anticipate all behaviors they might show as well as all the capabilities they might show. 
This is particularly relevant for discovering safety risks that such systems might expose. 
The systematic probing for such safety risks is known as red teaming.\autocite{Avin_2021, ganguli2022red} 
It is currently mostly performed by domain experts, even though there are some developments aiming to automate this process.\autocite{mazeika2024harmbench, hong2024curiositydriven}

However, systematic probing of models can also help discover capabilities that have not been discovered yet. 
Some, such as \textcite{lu2025automated} proposed to also automate this process by having a model compile a pool of interesting tasks. 

The advantage of red-teaming is that it can reveal insights that a benchmark cannot reveal (as a benchmark is constrained to measure one, or a few, capabilities). 
However, red teaming results are much more difficult to systematically compare and to scale.

\paragraph{Real-World Deployment Studies}

Evaluation through deployment in the intended use case is the most meaningful (in terms of measuring real-world impact) but the least standardized form of assessment. 
This could involve tracking how models influence decision-making in materials discovery campaigns or measuring their impact on experimental efficiency. 
While harder to quantify, such evaluations directly address the ultimate goal of practical utility.
An early example of this paradigm in materials science has been reported by \textit{toner2024artificial} who report that the \enquote{randomized introduction of a new materials discovery technology} in the R\&D lab of a materials company led to the discovery of 44\% more materials and a 39\% increase in patent filing.

\paragraph{Ablation Studies and Systematic (Stress) Testing}

Controlled experiments that systematically vary model components or test specific capabilities (like invariance preservation or physical constraint satisfaction) can provide mechanistic insights into model behavior. 

Such systematic tests are also important for understanding the robustness of models. 
One of the most common failure modes of \gls{ml} models in the real world is that the distribution they are tested on is different from the one they are trained on.\autocite{JMLR:v23:20-1335, hendrycks2021many} 
To assess the risk for this, it can be valuable to test the model under shifted distributions or to perform specific adversarial attacks.\autocite{Zhong_2022}  

For instance, \textcite{wang2024evaluating} added superfluous or distracting information to the prompts of \glspl{llm} and found that especially superfluous information can lead to a large performance degradation. 
We observed similar effects when testing multimodal models, where often the same information provided as text behaved differently from the information provided as image.\autocite{alampara2024probinglimitationsmultimodallanguage}
It is important to note that some of these attack vectors can also be used productively. 
For instance, \textcite{li2024outofdistribution} performed training in which they generated adversarial examples to increase robustness.

One example of a systematic stress test is the TEA Challenge 2023, a comprehensive evaluation of \glspl{mlff} for molecules, materials, and interfaces.\autocite{Poltavsky_2025} 
Model developers were given training datasets with limited information about the data generation process. 
They trained their models, which were then handed over to the organizers, who conducted \gls{md} simulations using the trained models --- attempting to asses the applicability and performance of different models in practical applications.

\paragraph{Comparative Head-to-Head Frameworks}  While traditional benchmarks evaluate models individually against fixed datasets, comparative frameworks directly pit models against each other. Similar to \gls{llm} evaluation platforms like Chatbot Arena,\autocite{chiang2024chatbot} these frameworks can implement pairwise comparisons where multiple models solve the same task, with relative rankings determined through direct comparison (e.g., via user feedback) rather than absolute metrics. This approach is particularly valuable for capturing qualitative aspects of model performance that may not be reflected in standard metrics, such as the usefulness of generated structures or the scientific plausibility of predictions. 
This paradigm, however, has not yet been used in materials science. 

\begin{practicalbox}[Practical Implications: Choosing Estimators]

Consider multiple evaluation approaches based on your goals:
  \begin{itemize}
  \item Traditional benchmarks: For standardized comparison
  \item Challenges/competitions: For prospective evaluation
  \item Red teaming: For discovering edge cases and vulnerabilities
  \item Real-world deployment: For ultimate validation of impact
  \item Ablation studies and stress tests: For mechanistic understanding and robustness assessment
  \end{itemize}
\end{practicalbox}

\section{Reporting an Estimate: The Critical Last Mile}

The final step in any measurement process is reporting the results. 
While this might seem straightforward, our experience with evaluation frameworks like ChemBench has shown that unclear or inconsistent reporting practices can severely limit the utility of evaluation results. 
Even the most carefully designed evaluation framework can fail to serve its purpose if it doesn't provide clear guidelines for how results should be reported and communicated.

Framework developers bear significant responsibility for ensuring consistent and meaningful reporting of results. 
This goes beyond simply specifying tasks and metrics --- it requires defining the complete pipeline from raw model outputs to final reported numbers. 
Some benchmarks, like MatBench,\autocite{Dunn_2020} handle this well by providing standardized reporting tools that generate structured JSON reports suitable for automated processing and comparison. 
In contrast, many recent \gls{llm} benchmarks only specify questions and answers without detailing crucial aspects like output parsing, handling of model refusals, or use of confidence scores. 
This lack of specificity renders reported numbers practically incomparable across different studies.

\paragraph{Beyond Summary Statistics }

While summary statistics enable quick comparisons, they often obscure important insights. 
For classification tasks, reporting full confusion matrices preserves information that might be lost in aggregate metrics. Similarly, for regression tasks, parity plots and error histograms provide crucial context beyond summary statistics like \gls{mae} or \gls{rmse}. 
This non-destructive reporting enables readers to gain deeper insights into model behavior. In bandgap prediction, for example, models often report only \gls{mae}, whereas parity plots reveal deviations that differ across the bandgap range --- information that remains hidden in aggregate metrics.\autocite{ottomano2024bandgap}

Similarly, claims about data efficiency should be supported by learning curves\autocite{huang2016communication} and where possible, plots of the performance against the similarity to training data points can give crucial insights into generalization performance.\autocite{jain2024deeplearning, ursu2024training, ektefaie2024evaluating}

\paragraph{Precise Metric Specification}

Even commonly used metrics require precise specifications. For instance, \enquote{AUC} could refer to different curves (\gls{roc}, \gls{pr}), and implementation details can affect results \autocite{blagec2020critical, Forman_2010}. 
Framework developers should provide exact mathematical definitions and reference implementations of all metrics. In catalyst adsorption studies, for example, reported \glspl{mae} can vary substantially depending on whether they are normalized per adsorbate, per atom, or per surface unit --- a choice often left unspecified.\autocite{Mancera2024}

\paragraph{Statistical Significance in Model Comparisons}  When reporting comparative results between models, statistical significance testing becomes crucial.\autocite{raschka2018model}
Seemingly superior performance may result from random variation rather than genuine model improvement. Standard practices should include reporting p-values from appropriate statistical tests (e.g., paired t-tests for models evaluated on the same test examples, or bootstrap tests for aggregate metrics), confidence intervals for performance differences, and effect sizes to quantify the magnitude of improvements.\autocite{ho2019moving}

\paragraph{Language Model Specific Challenges}

The evaluation of language models presents unique reporting challenges. 
Critical details include how model outputs are parsed, how refusals or invalid outputs are handled, and whether confidence scores or log probabilities are used. It is also crucial to specify exactly how the model has been used (prompting templates) and how sampling has been performed (e.g., greedy decoding, constrained decoding).
Without standardization of these aspects, reported results can become effectively incomparable.

\begin{practicalbox}[Practical Implications: Reporting Estimates]
\begin{itemize}
\item Standardize result reporting to ensure comparability:
  \begin{itemize}
  \item Use structured formats (e.g., JSON) for machine-readable results
  \item Provide reference implementations for metric calculations
  \item Document exact parsing/processing of model outputs
  \end{itemize}
\item Go beyond summary statistics:
  \begin{itemize}
  \item Include full confusion matrices for classification tasks
  \item Provide parity plots and error distributions for regression
  \item Report performance across meaningful data subgroups
  \end{itemize}
\item Specify metrics with complete precision:
  \begin{itemize}
  \item Document handling of edge cases (missing values, refusals)
  \item Include confidence intervals or uncertainty estimates
  \end{itemize}
\item For language models and agents:
  \begin{itemize}
  \item Detail prompting strategies and any post-processing
  \item Document handling of invalid outputs or refusals
  \item Report both average and distribution of performance
  \end{itemize}
\item Implement evaluation cards to document all aspects of measurement systematically
\end{itemize}
\end{practicalbox}

The next section introduces evaluation cards as a systematic approach to documenting these and other crucial aspects of evaluation frameworks.

\section{Evaluation Cards: Promoting Transparency in Model Assessment }

A recurring theme throughout this article has been the complexity and nuance involved in evaluating machine learning systems for materials science. 
Every evaluation framework involves numerous design choices and tradeoffs, from the selection of what to measure to the specific implementation details of how to measure it. 
These choices fundamentally shape what we can learn from an evaluation and how we should interpret its results. 
Given the critical role that evaluations play in scientific progress and decision-making, we need systematic ways to communicate about these choices and their implications.

Drawing inspiration from model cards \autocite{mitchell2018model} and data cards \autocite{pushkarna2022data}, we propose the development and adoption of evaluation cards. 
These structured documents aim to capture the key aspects of an evaluation framework through the lens of measurement theory while promoting transparency about design choices, limitations, and tradeoffs.

The structure of our proposed evaluation cards follows the estimand-estimator-estimate framework we used to structure this article, supplemented with practical sections needed for implementation and maintenance. 
A template can be found at \url{https://github.com/lamalab-org/eval-cards}. We also provide a HuggingFace space as gallery \url{https://huggingface.co/spaces/jablonkagroup/eval-cards-gallery}. 

The evaluation card template begins with sections capturing the motivation and design choices behind the evaluation approach. 
This includes explicit discussion of why particular evaluation strategies were chosen and what tradeoffs were considered. 
The estimand section then documents what is being measured, including where the measurement falls on the representational-pragmatic continuum and how it relates to real-world performance. 
The estimator section details the methodology, including metrics, technical requirements, and constraints. 
Finally, the estimate section specifies how results should be reported and reproduced.

Importantly, evaluation cards include dedicated sections for documenting known issues, limitations, and biases. 
This transparency is essential for adequately interpreting evaluation results and helps prevent misuse. 
The template also includes practical sections on versioning, maintenance, and usage guidelines to support the evaluation framework's evolution over time. \\

\noindent We envision evaluation cards serving multiple purposes:

\begin{itemize}
    \item For evaluation framework developers, they provide a structured way to think through and document design choices
    \item For users of evaluations, they offer a crucial context for interpreting results and understanding limitations
    \item For the broader community, they enable more informed comparison between different evaluation approaches
    \item For future researchers, they provide a historical record of how evaluation practices have evolved
\end{itemize}

\noindent As the field of materials science \gls{ml} continues to mature, we expect evaluation cards to evolve as well. 
Community feedback and practical experience will be essential in refining these templates to better serve the needs of the field.

We hope that by promoting the use of evaluation cards, we can move toward more transparent and rigorous evaluation practices in materials science machine learning. 
This transparency is essential for scientific progress and building trust in these increasingly important tools.

\section{Frontiers}
As ML becomes integral to materials science, several evaluation frontiers demand attention and close collaboration between experimental and computational communities. 

\subsection{Materials Science Specific Challenges}
\paragraph{Multi-Objective Metrics With Relation to Applications:} Materials typically require satisfying competing objectives simultaneously. For achieving real-world impact we must also aim to consider metrics that have a relation to the real world.\autocite{Sarewitz2016} As put by \textcite{Cheetham_2024}: \enquote{a compound can be called a material when it exhibits some functionality and, therefore, has potential utility.}

\paragraph{Uncertainty Quantification:} For many applications that are  high-risk or expensive, understanding prediction confidence is as important as the prediction itself. Materials-specific uncertainty methods must account for synthesis variability and application-specific risk tolerances.

\paragraph{Physical Consistency:} Materials models should respect fundamental physical laws. Developing systematic methods to evaluate physical consistency—beyond simple accuracy metrics—requires bridging \gls{ml} evaluation with physics-based validation.

\paragraph{Synthesizability Assessment:} A critical frontier is developing robust methods to evaluate whether computational predictions yield experimentally realizable materials, considering reaction pathways, kinetic barriers, and processing conditions typically absent from computational benchmarks. \\

\noindent These frontiers represent areas where materials science evaluation must transcend traditional \gls{ml} benchmarking to create assessment frameworks that meaningfully accelerate real-world materials discovery. Beyond this, there are also more general challenges. 

\subsection{General Challenges}

\paragraph{Understanding Data Generation Processes}
The \gls{ml} community often treats datasets as given, but understanding the data generation process is crucial for meaningful evaluation. 
This is especially relevant in materials science, where data might come from various sources such as experiments, simulations, or literature, each with their own biases and noise characteristics. 
A deeper understanding of data generation processes would enable more precise definitions of out-of-distribution generalization and guide the development of synthetic datasets for testing specific capabilities. 
Moreover, it would help identify potential failure modes and biases in our evaluation frameworks, leading to more robust assessment methods.

\paragraph{Formal Evaluation Theory}

While machine learning evaluation has largely developed through practical experience, there is growing potential for more formal theoretical approaches. 
Item response theory, widely used in educational testing, offers promising frameworks for understanding how different tasks probe model capabilities. 
Such formal approaches could help us better understand what our evaluations actually measure and how to design more informative evaluations.  
For instance, item response theory could help quantify how different material compositions or structure types contribute to task difficulty, leading to more balanced and informative test sets.

\paragraph{Quantifying Task Similarity}
One of the most pressing challenges is developing rigorous methods for measuring task similarity. 
This is particularly crucial for evaluating transfer learning capabilities, where understanding the relationship between source and target tasks could provide insights into model generalization. 
Current approaches often rely on informal or intuitive notions of task similarity, but we lack formal frameworks for quantifying these relationships. 
For example, predicting formation energy and bandgap might share underlying physics, but how do we quantify this relationship to predict transfer learning success? 
Developing meaningful measures of task similarity would not only improve our evaluation of transfer learning but could also guide the design of more efficient learning strategies.

\paragraph{Designing for Failure Modes}
Creating benchmarks that deliberately elicit interesting failure modes, rather than just showcasing model capabilities, represents another important frontier. 
This approach requires careful thought about what constitutes an \enquote{interesting} failure and how to systematically probe for vulnerabilities. 
In materials science, this might involve creating test cases that challenge physical consistency (like structures at the edge of stability), probe edge cases in composition space (such as rare earth elements), or evaluate robustness to experimental noise. 
For instance, a model might perform well on typical perovskite structures but fail on edge cases with unusual octahedral tilting. 
Understanding such failures often provides more insight than understanding successes.

\paragraph{Evaluation at Scale}
As models become more complex, ensuring efficient and practical evaluation becomes crucial. 
This challenge is particularly acute in materials science, where evaluating certain properties at high fidelity might require expensive computational simulations or laboratory experiments. 
We need to develop strategies for efficient evaluation that maintain rigor while being practically feasible. This might involve developing smart sampling strategies, leveraging cheaper proxy metrics where appropriate, or creating hierarchical evaluation frameworks that apply more expensive tests only to promising candidates.

\paragraph{LLMs as Evaluation Tools}
For highly complex tasks, particularly those involving reasoning or creative solutions, \glspl{llm} might become necessary tools for evaluation.\autocite{gu2024survey} 
This raises interesting questions about ensuring consistent and reliable judgments and validating \gls{llm}-based evaluations. 
We need to develop methods for combining automated and human evaluation while handling potential biases in \gls{llm} judgments.

\paragraph{Benchmark Maintenance and Evolution}

The current state of benchmark maintenance is concerning: many benchmarks in materials science \gls{ml} are effectively abandoned after publication.
Leaderboards become stale, code repositories unmaintained, and documentation outdated. This creates practical problems as dependencies break, \glspl{api} change, and our understanding of materials evolves.

A particularly problematic aspect is the continued use of effectively \enquote{solved} benchmarks. 
For example, the QM9 dataset, while historically important, has seen models achieve near-\gls{dft} accuracy, reaching the inherent noise limit of the underlying calculations. 
Yet, it continues to be used, potentially giving a false sense of progress when new models achieve marginal improvements on an already-solved problem.

The materials science community has yet to widely adopt basic software development practices like semantic versioning or automated testing. While funding often exists for creating new benchmarks, maintaining existing ones typically receives less attention, creating a perverse incentive to create new benchmarks rather than improve existing ones. A recent example for the importance of transparent communications by the developer suites is a report by \textcite{Lee_2016} that some (propietary) models had favorable rates of sampling and deprecating in the LLM Chatbot arena and hence, led to distorted scores.\\ 

\noindent We propose that benchmark developers should:

\begin{enumerate}
    \item Define clear criteria for benchmark deprecation at creation time
    \item Specify performance thresholds that would indicate the benchmark is \enquote{solved}
    \item Plan for successor benchmarks that test more challenging aspects of the same capabilities
    \item Maintain clear documentation about the benchmark's current status and limitations
    \item Actively communicate when a benchmark should no longer be used as a primary evaluation tool
\end{enumerate}

\noindent Technically, this requires 

\begin{enumerate}
    \item  Adoption of existing software development best practices like semantic versioning
    \item Implementation of automated testing and continuous integration
    \item Development of clear governance models and contribution guidelines
    \item Creation of sustainable funding models for long-term maintenance
    \item Establishment of standard protocols for benchmark deprecation and succession planning
\end{enumerate}

\paragraph{Meta-Evaluation Frameworks}
As evaluation methods become more sophisticated, we need better ways to evaluate the evaluations themselves. This includes developing metrics for assessing evaluation quality, methods for comparing different approaches, and frameworks for understanding biases and limitations. 
For example, how do we compare to evaluation frameworks that claim to measure the same underlying concept?

The advancement of these frontiers requires broad community engagement and collaboration between machine learning experts, materials scientists, and evaluation specialists. 
Progress in these areas would not only improve our ability to measure and guide progress in materials science \gls{ml} but could also inform evaluation practices in other scientific domains where similar challenges exist.

\section{Conclusions}

In \gls{ml}, especially in \gls{ml} for materials science, the measurement and the measurement instrument are often closely entangled. 
This is in contrast to (representational) measurements we are used to in the physical world. 
Some of this cannot easily be resolved as we aim to measure phenomena and behaviors that do not have an instantiation in the physical behavior. 

The entanglement of measurement and measurement instruments, however, implies that small changes to the instrument can have a notable impact on the measurement results. 

Commonly used \gls{ml} systems are underspecified.\autocite{JMLR:v23:20-1335} 
That is, many parameter choices are \enquote{hidden} and lead to equivalent performance under one less but potentially very different generalization performance. 
Those equivalent models are sometimes referred to as Rashomon sets, and it has been found that simpler models within those Rashomon sets tend to perform comparably to more complex models on noisier datasets.\autocite{semenova2023path} 
Using just one lens to understand model performance will not allow us to resolve the degeneracy in Rashomon sets. 
We rather need to embrace a vast diversity of different evaluation approaches to fully understand model performance and limitations.  
Thus, we should expand the toolbox of evaluation techniques we use in \gls{ml} for materials science.

Given the importance measurements have in driving progress in the field, transparency, open science, and craftsmanship in tool building are more important than ever. 
One might even argue for a dedicated science of building evaluations of \gls{ml} models. 

\printglossaries

\section*{Acknowledgements}
The work has been supported by the Carl-Zeiss Stiftung and OpenPhilanthropy. The work of M.S-W.\ was supported by Intel and Merck via the AWASES programme. K.M.J.\ is part of the NFDI consortium FAIRmat funded by the Deutsche Forschungsgemeinschaft (DFG, German Research Foundation) – project 460197019.

\section*{Conflicts of Interests}
K.M.J.\ has been a paid contractor for OpenAI (as part of the red teaming network).

\section*{Data availability}
No primary research data was created or used in this work.

\section*{Declaration of generative AI and AI-assisted technologies in the writing process}

During the preparation of this work the authors used Anthropic's Claude models in order to improve language and readability. After using this service, the authors reviewed and edited the content as needed and take full responsibility for the content of the publication.

\printbibliography 

\end{document}